\documentclass[american,manuscript]{revtex4-1}
\usepackage[T1]{fontenc}
\usepackage[utf8]{inputenc}
\usepackage{color}
\usepackage{babel}
\usepackage{array}
\usepackage{multirow}
\usepackage{amsmath}
\usepackage{graphicx}
\usepackage{setspace}
\usepackage[unicode=true]
 {hyperref}

\makeatletter

\providecommand{\tabularnewline}{\\}
\newcommand{\lyxdot}{.}

\makeatother

\begin{document}

\title{Phase-field modeling of Li-insertion kinetics in single LiFePO\textsubscript{4}-nano-particles for
rechargeable Li-ion battery application }

\author{Michael Fleck}
\email{michael.fleck@uni-bayreuth.de}

\affiliation{Metals and Alloys, University Bayreuth, Ludwig-Thoma-Straße 36b, 95447 Bayreuth, Germany}

\author{Holger Federmann}

\affiliation{TenneT TSO GmbH, Bernecker Straße 70, 95448 Bayreuth, Germany}

\author{Evgeny Pogorelov}

\affiliation{Advanced Ceramics Group, University Bremen, Bibliothekstraße 1, 28359 Bremen, Germany}

\preprint{{\tiny{}\copyright~ 2018. This version of the manuscript is made available under the CC-BY-NC-ND 4.0
license \href{http://creativecommons.org/licenses/by-nc-nd/4.0/}{http://creativecommons.org/licenses/by-nc-nd/4.0/}
}\includegraphics[width=1cm]{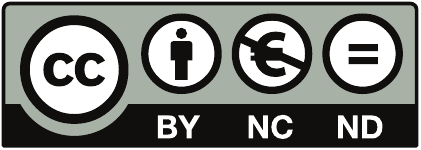}}
\begin{abstract}
\begin{singlespace}
We develop a continuum phase-field model for the simulation of diffusion limited solid-solid phase transformations
during lithium insertion in LiFePO\textsubscript{4}-nano-particles. The solid-solid phase boundary between
the LiFePO\textsubscript{4} (LFP)-phase and the FePO\textsubscript{4} (FP)-phase is modeled as a diffuse
interface of finite width. The model-description explicitly resolves a single LiFePO\textsubscript{4}-particle,
which is embedded in an elastically soft electrolyte-phase. Furthermore, we explicitly include anisotropic
(orthorhombic) and inhomogeneous elastic effects, resulting from the coherency strain, as well as anisotropic
(1D) Li-diffusion inside the nano-particle. In contrast to other related research work, we employ an
Allen-Cahn-type phase-field approach for the diffuse interface modeling of the solid-solid phase boundary.
The model contains an extra non-conserved order parameter field to distinguish the two different phases.
The evolution of this order parameter field is controlled by an extra kinetic parameter independent from
the Li-diffusion.  Further, the effect of the nano-particle's size on the kinetics of FP to LFP phase
transformations is investigated by means of both model. Both models predict a substantial increase in
the steady state transformation velocity as the particle-size decreases down to dimensions that are comparable
with the width of the interface between the FP and the LFP-phase.  However, the extra kinetic parameter
of the Allen-Cahn-type description may be used to reduce the strength of the velocity-increase with the
decreasing particle size.  Further, we consider the influence of anisotropic and inhomogeneous elasticity
on the lithiation-kinetics within a rectangularly shaped LiFePO\textsubscript{4}-particle embedded in
an elastically soft electrolyte. Finally, the simulation of equilibrium shapes of LiFePO\textsubscript{4}-particles
is discussed. Within a respective feasibility study, we demonstrate that also the simulation of strongly
anisotropic particles with aspect ratios up to 1/5 is possible.\\
\vspace{0.5cm}
The original research article is available at: \href{https://doi.org/10.1016/j.commatsci.2018.06.049}{https://doi.org/10.1016/j.commatsci.2018.06.049}
\end{singlespace}
\end{abstract}
\maketitle

\section{Introduction}

\label{Introduction}LiFePO\textsubscript{4} powder is widely considered to be a promising novel cathode
material for the application in rechargeable Li-ion batteries. The reasons are high energy storage, low
cost and electrochemical stability \cite{PadhiNanjundaswamy1997,Xu201251}. Recent trends in the design
of LiFePO\textsubscript{4}-cathodes is to synthesize particles of smaller and smaller sizes. For LiFePO\textsubscript{4}-nano-particles
excellent performance parameters such as high charge rates have been reported \cite{KangCeder052009}.
However, designing next generation Li-ion battery cathodes materials, based on LiFePO\textsubscript{4}nano-particles,
a thorough understanding of the kinetics of the lithiation process in this novel battery material is
highly desired. Here, we aim to contribute to this topic from the perspective of a continuum materials
science simulation approach. Within the present article we focus on the issues that are placed around
the model development, whereas subsequent detailed simulation studies and related results will be part
of future work.

The true physical mechanism of the Li insertion process in cathodes made of LiFePO\textsubscript{4}
nano-powders is in due to the underlying complexity still a matter of an ongoing scientific debate, and
for recent reviews on this issue, we refer to \cite{Malik01012013,Love01012013}. In this article, the
charge and discharge process in the bulk LiFePO\textsubscript{4}-material is regarded to proceed via
a coherent solid-solid phase transformation between LiFePO\textsubscript{4} (LFP-phase) and FePO\textsubscript{4}
(FP-phase). During this first order phase transformation lithium is inserted into the olivine bulk material.
The diffusion of Li inside the orthorhombic olivine lattice is strongly anisotropic along 1D channels
in (010) crystallographic direction \cite{TangCarterChiang082010,NishimuKobayasOhoyama092008,MalikBurchBazantCeder2012}.
Important for the modeling of the Li insertion process in single LiFePO\textsubscript{4}-nano-particles
is of course the thermodynamics of the bulk LiFePO\textsubscript{4}-system: At room temperature the
two stable phases are separated by a miscibility gap, generally providing a strong tendency for the material
to phase separate into FP and LFP phase. It is quite interesting that for small LiFePO\textsubscript{4}-particles,
a size dependence of the miscibility gap has been observed experimentally \cite{MeethonHuangCarter032007,GibotCasas-CLaffont092008,WagemakerSinghBorgholsLafontHaverkatePeterson2011},
whereas a relation to the excellent rate capabilities of nano-sized LiFePO\textsubscript{4}-powders
is conceivable. It is interesting to note that, such a size dependence of the miscibility gap can result
from a gradient energy contribution within diffuse interphase models, where the interface width is on
the same order of magnitude than the size of the nano-particle \cite{BurchBazant2009,CogswellBazant2013,PogorelovKundinFleck2017}. 

For the development of a respective diffuse-interface description, we apply the so-called phase-field
method. This approach has already been applied to model electrochemical reaction kinetics in electrode
materials in a number of other research work \cite{TangBelakDorr032011,CogswellBazant2012,CogswellBazant2013,Orvananos01012015,LiangQiXue112012,Guyer20042,HuLiRossoSushko2013}.
Within the phase-field method, moving phase boundaries between different phases are treated as diffuse
interfaces of finite width. Then, the evolution of the diffuse phase boundary is driven by the mechanics
and the thermodynamics of the adjacent bulk phases. In turn, the motion of the diffuse interface strongly
influences the bulk properties such as mechanical or thermodynamical degrees of freedom. Already on microscopic
length-scale problems, such as solidification or solid-state reactions in metallic alloys, the diffuse
interface approach provides an elegant way to dynamically incorporate complex effects, such as multi-component
diffusion of refractory elements, chemical reactions at the phase boundary or stress and strain effects
due to the lattice mismatch between the phases \cite{AstaBeckermKarma012009,Emmeric012008,Steinba2009,WangLi012010,FleckQuerfurthGlatzel2017,MushongeraFleckKundinWangEmmerich2015}.
On the nano-scale, such as in novel LiFePO\textsubscript{4} cathode materials, the width of the diffuse
interface can be chosen in accordance with corresponding experimental observations. In contrast to phase-field
descriptions on the micrometer scale, then, the diffuse interface of finite width can carry physical
information, which means that the respective nonlinear model behavior has an actual physical meaning
\cite{BurchBazant2009,WagemakerSinghBorgholsLafontHaverkatePeterson2011,CogswellBazant2013,PogorelovKundinFleck2017,TangKarma062012}. 

Further, we also include anisotropic and inhomogeneous elastic effects, resulting from the lattice misfit
between the LFP and the FP-phase. Primarily, these effects are considered to be very important to reproduce
experimental observations with regard to favored LFP/FP-interface orientations \cite{VanderVenGarikiKim2009,CogswellBazant2012}.
Furthermore, it is also interesting to access the micro-mechanical states during the charging of LiFePO\textsubscript{4}-particles,
since frequent cracking along the FP/LFP-interface upon electro-chemical shock is observed and discussed
\cite{Zhu01012012,WoodfordCarterChiang2012,WoodfordChiangCarter2010}. However, a realistic consideration
of the elastic effects requires the modeling of the whole particle, which can have\emph{ }strongly anisotropic
shapes \cite{ChenSongRichard042006,QuiWangXieLiWenWang2012}. 

Therefore, here, we develop a phase-field model, which combines the following mechanisms for a description
of the lithiation-reaction in single LiFePO\textsubscript{4}-particles: 
\begin{itemize}
\item The introduction of anisotropic bulk diffusion along 1D channels in (010) crystallographic direction 
\item The incorporation of anisotropic coherency strains arising from the lattice-mismatch between the two
joining solid phases with different elastic constants. 
\item The implementation of strongly anisotropic interfacial energies that give rise to the anisotropic particle-shapes 
\item The introduction of the anisotropic particle-shape, which act as a free-surface guaranteeing realistic
strain energy contributions 
\end{itemize}
A major difference of the present model, as compared to other models of similar purpose is that the diffusion
limited phase transformations are reformulated in terms of two strongly coupled but still independent
kinetic equations: one for the Li diffusion and one for the solid-solid phase transformations. This leads
to two coupled 2nd order partial differential equations of parabolic type, instead of one equation of
forth order. Within the present article, we focus on the model development and discuss the relations
and differences to other phase field models for the simulation of charge and discharge in LiFePO\textsubscript{4}
cathodes. Moreover, first interesting results on the size dependent kinetics of diffusion limited phase
transformations are presented. Since, an electrolyte phase is explicitly included, it might be also possible
to study the complex dynamics of multi-particle interaction \cite{Dreyer20111008,DreyerJamnikGuhlke042010,Orvananos01012014,Orvananos01012015}.
However, this is beyond the scope of the present work.

The article is structured as follows: In section \ref{sec:Phase-field-model} the development of the
phase-field model for the kinetic simulation of Li-insertion in single LiFePO\textsubscript{4} particles
is described. It is subdivided into the description of the energetics and the subsequent derivation of
the evolution equations. Then, in section \ref{sec:Simlation-results}, the simulation results are presented.
First, we present the results on the size-dependent kinetics of diffusion limited phase transformations,
second the influence of anisotropic and inhomogeneous elastic effects is discussed and finally a feasibility
study on the formation of strongly anisotropic particle shapes is presented. Finally, in section \ref{sec:summary},
a small summary is given.

\section{Phase Field Modeling of LiFePO\textsubscript{4} Particles }

\label{sec:Phase-field-model} 
\begin{figure}[h]
\centering{}\includegraphics[width=0.7\linewidth]{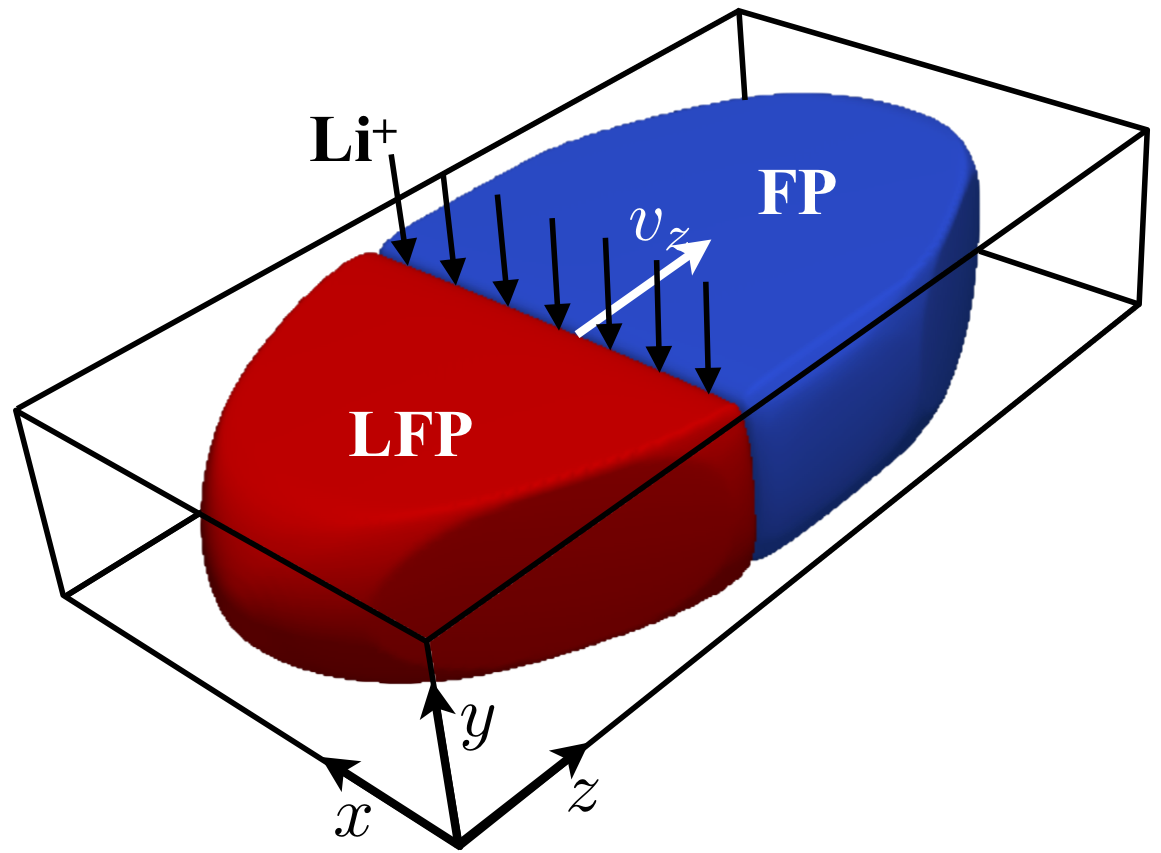} \caption{\label{fig:par_3D} (color online) The setup of a partially lithiated, plate-like particle for the subsequent
3D phase field simulation of the lithiation-reaction inside single LiFePO\textsubscript{4} nano-particles.
The LiFePO\textsubscript{4}-phase is shown in red and the FePO\textsubscript{4}-phase in blue. The
particle is surrounded by an electrolyte phase, not shown in the picture. Li-diffusion inside the particle
is restricted to one-dimensional channels oriented along the (010)-crystallographic orientation as indicated
by black arrows.}
\end{figure}

Here, we describe the development of the continuum phase field model
for the simulation of lithium insertion kinetics in single LiFePO\textsubscript{4} nano-particles. In
contrast to conventional Cahn-Hilliard-type descriptions, the lithium transport is considered to be independent
from the solid-solid phase transformation from the FP-phase to the LFP-phase. The conservative lithium
transport is described by a continuous concentration field $c$. In order to locally distinguish between
the LFP and the FP-phase, we introduce the non-conserved phase field parameter $\phi$, with $\phi=1$
denoting the LFP phase and $\phi=0$ denoting the FP-phase. At the FP/LFP-interface a smooth transition
about the phase field width $\xi_{i}$ between the two bulk values of the phase field $\phi$ is enforced.
Furthermore, in the model, we resolve individual nano-particles with the potential possibility to study
multi-particle interaction \cite{DreyerJamnikGuhlke042010,Dreyer20111008,LiEl-GabaFerguso122014,Orvananos01012014,Orvananos01012015}.
This is realized by the inclusion of an extra phase field $\varphi$ for the particle, which takes the
value $\varphi=1$ at the places where a LiFePO\textsubscript{4}-particle is located and the value $\varphi=0$
in the area that surrounds the particles. Finally. the elastic deformations, which result from the lattice-misfit
between two the coherently connected phases are contained in the displacement vector-field $\mathbf{u}$.
All these field are defined in the entire (rectangular) simulation domain. The full 3D-setup of the model
for the simulation of lithium insertion in a single anisotropically shaped LiFePO\textsubscript{4}-particle
is illustrated in Fig.~\ref{fig:par_3D}. 

\subsection{The energetic formulation}

Now we write down the actual phase-field model consisting of the description of the LFP/FP interface,
described by $\phi$, and the particle-phase boundary, which is denoted as $\varphi$. We start from
a phenomenologically motivated Helmholtz free energy functional 
\begin{align}
F & =\int dV\Bigg(f_{c}+f_{e}+f_{i}+f_{p}\Bigg).\label{eq:Free-energy-functional}
\end{align}
The first two terms account for thermodynamics and mechanics of the three bulk phases. $f_{c}(\phi,\varphi,c)$
denotes the chemical free energy density, 
\begin{align}
f_{c} & =\bar{f}_{0}+\frac{\bar{X}}{2}\left(c-\bar{c}_{eq}\right)^{2},\label{eq:chemical-free-energy-density}
\end{align}
which is a quadratic expansion of the free energy density with respect to the local concentration $c(\mathbf{x},t)$
of Li$^{+}$-ions, with $\bar{f}_{0}(\phi,\varphi)$ being the phase dependent free energy minimum, $\bar{X}(\phi,\varphi)$
denotes the phase dependent thermodynamic factor, and $\bar{c}_{eq}(\phi,\varphi)$ denotes the phase
dependent concentration of minimal free energy. The second contribution is the elastic free energy density,
\begin{align}
f_{e} & =\frac{1}{2}\left(\epsilon_{ik}-\overline{\epsilon}_{ik}^{0}\right)\overline{C}_{iklm}\left(\epsilon_{lm}-\overline{\epsilon}_{lm}^{0}\right),\label{eq:elastic-free-energy-density}
\end{align}
where $\epsilon_{ik}=\left(\partial u_{k}/\partial x_{i}+\partial u_{i}/\partial x_{k}\right)/2$ denotes
the symmetric strain tensor, $\mathbf{u}(\mathbf{x},t)$ denotes the displacement vector-field, $\overline{\epsilon}_{ik}^{0}(\phi,\varphi)$
is the phase dependent symmetric transformation strain tensor (eigenstrain), and $\overline{C}_{iklm}(\phi,\varphi)$
denotes the phase dependent stiffness tensor. Note that, according to Einstein's sum convention, the
summation over repeated indices is implicitly defined. For any of the mentioned phase dependent physical
parameters $\bar{P}=\bar{X},\bar{c}_{eq},\bar{f}_{0},\bar{\epsilon}_{ik}^{0},\bar{C}_{iklm}$ the phase
field dependence is always of the following type, $\bar{P}=h(\phi)h(\varphi)P^{LFP}+h(1-\phi)h(\varphi)P^{FP}+h(1-\varphi)P^{E}$,
where $P^{LFP/FP/E}$ denote respectively the particular bulk value of the specific parameter in the
LFP phase, the FP phase and the surrounding electrolyte phase, and $h(\phi)=\phi^{2}(3-2\phi)$ is the
interpolation function. This interpolation function corresponds to the minimal polynomial function satisfying
the necessary interpolation condition, $h(0)=0$, $h(1)=1$, and having also vanishing slope at zero
and one, i.e.~$\partial h(\varphi=0,1)/\partial\varphi=0$, for not to shift the two minima in case
of finite driving forces \cite{Kassner2001}.

The last two contributions to the free energy functional are respectively related to the LFP/FP interface
energy and to the surface energy of the particle. 
\begin{align}
f_{i} & =\frac{3\gamma_{i}\xi_{i}}{2}\left(\nabla\phi\right)^{2}+\frac{2\gamma_{i}}{\xi_{i}}g\left(\phi\right),\label{eq:free-energy-density-of-LFP-FP-interface}
\end{align}
where $g(\phi)=\phi^{2}(1-\phi)^{2}$ is the double-well-potential, and $\gamma_{i}$ as well as $\xi_{i}$
denote the interfacial energy as well as the phase field width for the FP to LFP solid solid transformation.
\begin{align}
f_{p} & =\gamma_{p}\left(\nabla\varphi\right)\left(\frac{3\xi_{p}}{2}\left(\nabla\varphi\right)^{2}+\frac{2}{\xi_{p}}g\left(\varphi\right)\right),\label{eq:free-energy-density-of-particle}
\end{align}
where  $\gamma_{p}\left(\nabla\varphi\right)$ is the orientation dependent surface energy of the anisotropic
particle and similarly $\xi_{p}$ denotes the phase field width of the particles surface.  Note that
the present anisotropic formulation, where the phase-field-gradient dependence is attributed to $\gamma_{p}\left(\nabla\varphi\right)$,
produces an orientational anisotropy of the interface energy, whereas the phase field width remains isotropic
\cite{FleckMushongPilipen102011}.

The functional dependence between the interfacial energy and the orientation is decomposed into a sum
of dimensionless facet-functions, which provide a sharp minimum for the respective facet orientation.
Therefore the full anisotropy profile is given by the sum over all facet functions: 
\begin{align}
\frac{\gamma_{p}(\nabla\varphi)}{\gamma_{0}} & =B+\sum_{\mu}\delta_{\mu}\alpha_{\mu}(\nabla\varphi)+\sum_{\nu}\delta_{\nu}\beta_{\nu}(\nabla\varphi),\label{eq:surface-energie}
\end{align}
where $B$ denotes the isotropic background and $\delta_{\mu/\nu}$ denote respective facet-strengths.
Here, two different kinds of functions, $\alpha_{\mu}(\nabla\varphi)$ and $\beta_{\mu}(\nabla\varphi)$
are introduced to interpolate the interfacial energy profile in the vicinity of a certain facet orientation.
The first anisotropy function is given by 
\begin{align}
\alpha_{\mu}(\nabla\varphi) & =-\left(\frac{\left(n_{l}^{\mu}\partial_{l}\varphi\right)^{2}}{\left(\nabla\varphi\right)^{2}}\right)^{d},\label{eq:Ani-Function-alpha}
\end{align}
with $\vec{n}^{\nu}$ being the unit vector normal to the plane of the $\nu-$th facet and $d=128$ is
the exponent which allows the orientational minimum to be strongly localized. This function is a generalization
of the facet function proposed in \cite{UeharaSekerka062003}. The second anisotropy function is a 3D-generalization
of the physically motivated anisotropy profile of Debierr et al.~\cite{DebierrKarmaCelesti102003}.
It is a stepwise defined function, 
\begin{align}
\beta_{\nu}(\nabla\varphi)= & \begin{cases}
\left(1-\cos\theta_{0}\:\cos\vartheta^{\nu}\right)/\sin\theta_{0} & \mathbf{if}\quad\vartheta^{\nu}>\theta_{0},\\
\left(1+\cos\theta_{0}\:\cos\vartheta^{\nu}\right)/\sin\theta_{0} & \mathbf{if}\quad\vartheta^{\nu}<-\theta_{0}\\
\left|\vec{n}^{\nu}\times\vec{\nabla}\phi\right|/\left|\nabla\phi\right| & \mathbf{else},
\end{cases},\label{eq:Ani-Function-beta}
\end{align}
where $\cos\vartheta^{\nu}=\vec{n}^{\nu}\cdot\vec{\nabla}\varphi/\left|\nabla\varphi\right|$ with $\vec{n}^{\nu}$
being the unit vector normal to the plane of the $\nu-$th facet. Here, $\theta_{0}$ denotes the smoothing
angle in order to overcome the cusp in the anisotropy profile at the facet orientation $\vartheta^{\nu}$. 

\subsection{The evolution equations}

\label{subsec:The-evolution-equations}With respect to the modeling of the Li-insertion into single LiFePO\textsubscript{4}-nano-particle
the conjointly and coupled motion of the phase fields is not required, as long as the particle morphology
does not change during the lithiation, which we will assume in the following. Therefore, we divide the
simulation into a sequence of two independent stages: First the formation of the anisotropic particle,
and second the lithiation of the finished particle, by means of a transformation from FP to LFP-phase.

The kinetics of the particle formation result from the corresponding Allen-Cahn-equation of motion 
\begin{align}
\frac{3\gamma_{p}\xi_{p}}{K_{p}}\frac{\partial\varphi}{\partial t} & =-\left(\frac{\delta F}{\delta\varphi}\right)\nonumber \\
 & =\nabla\left(\frac{\partial f_{p}}{\partial\left(\nabla\varphi\right)}\right)-\frac{\partial f_{p}}{\partial\varphi}-\frac{\partial f_{c}}{\partial\varphi},\label{eq:PFM-Phase-field-eq-anisotropic}
\end{align}
where $K_{p}$ being the kinetic coefficient for morphological changes of the particles shape. Note that
the elastic contribution is omitted here, because we do not consider elastic effects during the formation
of the particle. To be able to relax to the equilibrium shape of the particle, we require the conservation
of the particle volume. Therefore, we consider the bulk free energy density of the electrolyte phase
$f_{0}^{E}$ to be time-dependent in such a way that a volume change of the electrolyte phase $V^{E}$
is prohibited \cite{FleckMushongPilipen102011,KarmaRappel041998}  
\begin{align}
0=\frac{d}{dt}V^{E}(t) & =\int_{V}\frac{\partial}{\partial t}\left(1-\varphi(\mathbf{x},t)\right)dV.\label{eq:PFM-volume-preservation}
\end{align}
Then, inserting the phase field equation~(\ref{eq:PFM-Phase-field-eq-anisotropic}), the time-dependency
of the constant chemical potential density of the electrolyte phase can be calculated as ${\color{blue}f_{0}^{E}(t)}=R(t)/H(t)$,
where the following abbreviations $R(t)=K\int_{V}\left(\delta F/\delta\varphi\right)dV$ and $H(t)=\int_{V}h'(\varphi)dV$
have been introduced. As shown in \cite{NestlerWendlerSelzer072008}, this method is also suited for
the multi-phase application.

For the second stage, the phase field of the particle $\varphi$  is treated as constant in time over
the whole subsequent simulation of Li-insertion by means of an FP to LFP phase transformation. However,
now addressing the FP to LFP solid-state reaction, elastic effects as well as the anisotropic Li-transport
both become very important. The Li$^{+}$-ion transport inside a nano-particle is considered to be the
rate limiting process, providing the slowest time-scale. The evolution of the local Li$^{+}$-ion concentration
is given by the continuity equation 
\begin{align}
\frac{\partial c}{\partial t} & =-\frac{\partial}{\partial x_{i}}\left(M_{ik}\frac{\partial\mu}{\partial x_{k}}\right),\label{eq:Li-Diffusion-eq}
\end{align}
where the ion transport inside the nano-particle is strongly anisotropic, which is reflected by the
tensorial mobility $\mathbf{M}$. Here, $\mu$ denotes the non-equilibrium chemical potential, which
can be derived from the free energy functional (\ref{eq:Free-energy-functional}) by means of the functional
derivative with respect to the concentration, 
\begin{align}
\mu & =-\frac{\delta F}{\delta c}=-\bar{X}\left(c-\bar{c}_{eq}\right).\label{eq:Definition-chemical-Potential}
\end{align}
For the seek of simplicity, we consider the lithium mobility tensor $\mathbf{M}$ as well as the thermodynamic
factor $X$ to be a phase independent here. Then from the continuity equation (\ref{eq:Li-Diffusion-eq})
as well as Eq.~(\ref{eq:Definition-chemical-Potential}) we can derive an equation of motion for the
non-equilibrium chemical potential $\mu$. We obtain, 
\begin{align}
\frac{\partial\mu}{\partial t} & =\frac{\partial}{\partial x_{i}}\left(D_{ik}\frac{\partial\mu}{\partial x_{k}}\right)-X\Delta c_{eq}\frac{\partial h}{\partial\phi}\frac{\partial\phi}{\partial t},\label{eq:chemical-potential-diffusion-eq}
\end{align}
where $\Delta c_{eq}=c_{eq}^{LFP}-c_{eq}^{FP}$ and $D_{ik}=XM_{ik}$ denotes the lithium diffusivity
tensor. The Li-ion transport outside the particle is assumed to be infinity fast leading to a constant
ionic concentration distribution given by the externally applied chemical potential. 

The elastic effects result from the lattice-misfit between the LFP and FP phase, which are in the order
of a few percent. Restricting to a purely elastic description, we assume the LFP/FP-interface to be coherent.
Then, the phase field concept naturally provides the mechanical equilibrium condition: The functional
derivative of the free energy functional (\ref{eq:Free-energy-functional}) with respect to the displacements
has to vanish, 
\begin{align}
0 & =-\frac{\delta F}{\delta u_{i}}=\frac{\partial}{\partial x_{k}}\left(\frac{\partial f_{e}}{\partial\epsilon_{ik}}\right)=\frac{\partial\sigma_{ik}}{\partial x_{k}}.\label{eq:Intro-Phasefield-mechanical-equilibrium}
\end{align}
The stress tensor is naturally defined as the derivative of the elastic free energy density with respect
to the strains, $\sigma_{ik}=\partial f_{e}/\partial\epsilon_{ik}$. The electrolyte phase is considered
to be elastically week i.e.~$C_{iklm}^{E}=0$, leading to a stress free surface of the particle.

Finally using variational principles, we obtain the following non-linear partial differential equations
for the evolution of the phase field describing the FP to LFP transformations,

\begin{align}
\frac{1}{K_{i}}\frac{\partial\phi}{\partial t} & =\nabla^{2}\phi-\frac{2}{\xi_{i}^{2}}g'(\phi)\nonumber \\
 & \quad-\frac{1}{3\gamma_{i}\xi_{i}}\left(\frac{\partial f_{c}}{\partial\phi}+\frac{\partial f_{e}}{\partial\phi}\right),\label{eq:Phase-field-eq-2D-sim}
\end{align}
where $K_{i}$ is the kinetic coefficient for the solid solid phase transformations.

In summary, the model consists of a set of coupled partial differential equations of second order, as
given by the Eq.~(\ref{eq:PFM-Phase-field-eq-anisotropic}) for the phase field of the particle, the
Eq.~(\ref{eq:chemical-potential-diffusion-eq}) for the Li diffusion, the Eq.~(\ref{eq:Intro-Phasefield-mechanical-equilibrium})
for the elastic displacements and Eq.~(\ref{eq:Phase-field-eq-2D-sim}) for the FP to LFP transformations
inside the particle. All these equations are solved by finite difference schemes operating on one fixed
square grid with explicit Euler-type time integration. With regard to the mechanical (elastic) equilibrium
Eq.~(\ref{eq:Intro-Phasefield-mechanical-equilibrium}), we perform a Jacobi relaxation. The underlying
finite difference scheme is formulated on a staggered grid, as has been further explained in \cite{FleckMushongPilipen102011,PilipenkoFleckEmmerich2011,Fleck012011}.

\subsection{Cahn-Hilliard formulation of FP to LFP transformations}

\begin{figure}
\begin{centering}
a)\includegraphics[width=0.45\linewidth]{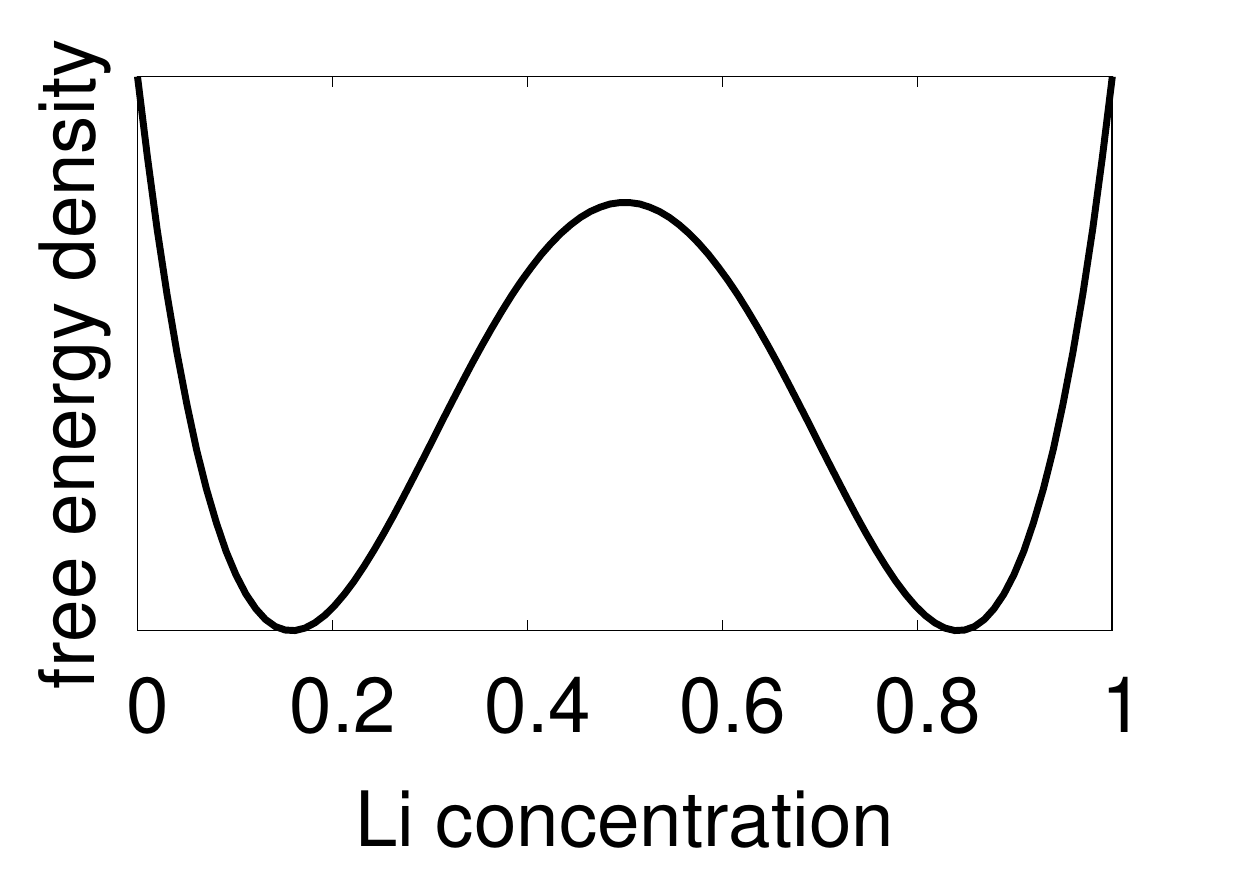} b)\includegraphics[width=0.48\linewidth]{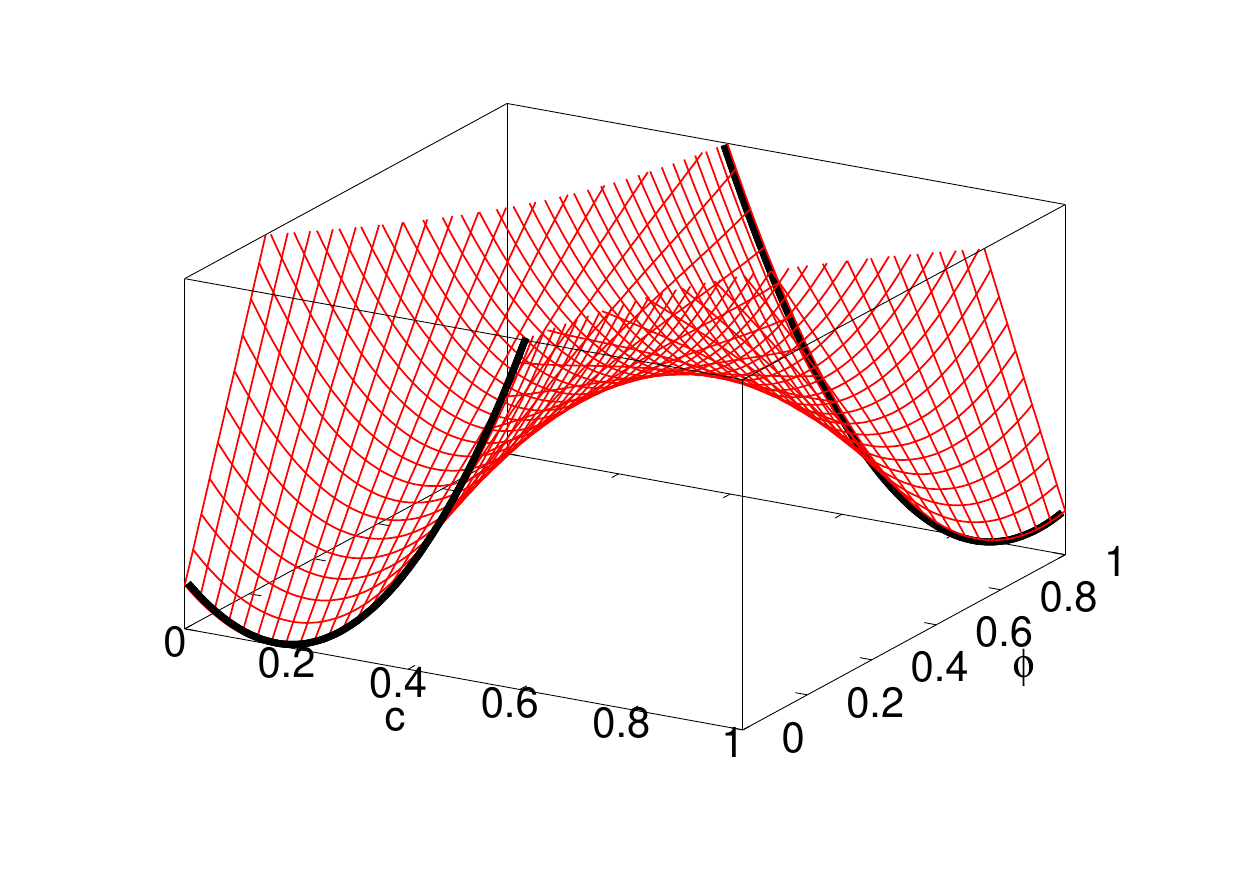}
\par\end{centering}
\caption{\label{fig:Comparison-of-the-energetics}Comparison of the energetic landscape of the two different models:
a) the Cahn-Hilliard-type model and b) the Allen-Cahn-type model.}
\end{figure}

In order to discuss differences and relations between the modeling of the FP to LFP transformations using
our new Allen-Cahn-model and conventional Cahn-Hilliard-type formulations, we also consider a respective
Cahn-Hilliard-type model. The equivalent Cahn-Hilliard-type formulation of the functional looks as follows
\begin{align}
F_{CH} & =\int dV\left(\frac{U}{2}\left(\nabla c\right)^{2}+f_{c}^{CH}(c)\right)\label{eq:CH-Functional}
\end{align}
where $f_{c}^{CH}=X^{CH}\left(c-c_{eq}^{LFP}\right)^{2}\left(c-c_{eq}^{FP}\right)^{2}/2$ and $U$ is
the gradient energy density. Here, $X^{CH}$ denotes the thermodynamic factor of the Cahn-Hilliard formulation.
The main difference in the energetics is that for the Cahn-Hilliard formulation it is a one dimensional
function of the concentration alone, whereas for the Allen-Cahn model it is a two dimensional function
of the concentration and the phase-field, as also illustrated in Fig.~\ref{fig:Comparison-of-the-energetics}.

Using Eq.~(\ref{eq:Li-Diffusion-eq}) as well as the variational definition of the chemical potential
Eq.~(\ref{eq:Definition-chemical-Potential}), we obtain the Cahn-Hilliard equation
\begin{align}
\frac{\partial c}{\partial t} & =M_{yy}\frac{\partial^{2}}{\partial x^{2}}\left(\frac{\delta F}{\delta c}\right),\nonumber \\
 & =M_{yy}\frac{\partial^{2}}{\partial x^{2}}\left(\frac{\partial f_{c}^{CH}}{\partial c}-U\nabla^{2}c\right).\label{eq:CH-equation}
\end{align}
The one dimensional equilibrium concentration profile of the Cahn-Hilliard model is given by 
\begin{align}
c_{0}(x,t) & =\frac{c_{eq}^{m}}{2}\left(1-\frac{\Delta c_{eq}}{c_{eq}^{m}}\tanh\left(\frac{x-x_{0}}{\xi_{i}}\right)\right),\label{eq:CH-eq-profile}
\end{align}
with $c_{eq}^{m}=c_{eq}^{LFP}+c_{eq}^{FP}$ and $\Delta c_{eq}=c_{eq}^{LFP}-c_{eq}^{FP}$. Using this
solution, we match the parameters $U,X^{CH}$ with the respective interface energy $\gamma_{i}$ from
the Allen-Cahn formulation presented in section \ref{subsec:The-evolution-equations}. Inserting the
equilibrium solution (\ref{eq:CH-eq-profile}) into the free energy functional (\ref{eq:CH-Functional}),
we can evaluate the total free energy density of the LFP/FP interface in the model,
\begin{align*}
\gamma_{i} & =F_{CH}[c_{0}(x,t)]\\
 & =X^{CH}\int_{-\infty}^{+\infty}dx\left(c_{0}-c_{eq}^{FP}\right)^{2}\left(c_{0}-c_{eq}^{LFP}\right)^{2}\\
 & =\frac{\xi_{i}\Delta c_{eq}X^{CH}}{12}\left(2\left(\left(c_{eq}^{FP}\right)^{3}-\left(c_{eq}^{LFP}\right)^{3}\right)+3\left(\left(c_{eq}^{FP}\right)^{2}+\left(c_{eq}^{LFP}\right)^{2}\right)\Delta c_{eq}\right),
\end{align*}
where we used that $\partial c_{0}/\partial x=2\left(c_{0}-c_{eq}^{FP}\right)\left(c_{0}-c_{eq}^{LFP}\right)/(\xi_{i}\Delta c_{eq})$.
 From this result, we obtain the condition
\begin{align}
X^{CH} & =\frac{1}{\xi_{i}\Delta c_{eq}}\frac{12\gamma_{i}}{2\left(\left(c_{eq}^{FP}\right)^{3}-\left(c_{eq}^{LFP}\right)^{3}\right)+3\left(\left(c_{eq}^{FP}\right)^{2}+\left(c_{eq}^{LFP}\right)^{2}\right)\Delta c_{eq}}.\label{eq:CH-energy-matching}
\end{align}
Further, inserting the equilibrium solution Eq.~(\ref{eq:CH-eq-profile}) into the Cahn-Hilliard equation
(\ref{eq:CH-equation}), we obtain $U=\xi_{i}^{2}\Delta c_{eq}^{2}X^{CH}/4.$

\section{Simulation results}

\label{sec:Simlation-results}

\subsection{Diffusion limited FP to LFP phase transformation wave}

First, we consider diffusion-limited FP to LFP phase transformations in a rectangular 2D domain, neglecting
elastic effects as well as the explicit representation of the electrolyte phase. The relevance of this
reduced configuration with regard to the behavior of LiFePO\textsubscript{4}-powders as cathode-material
has been discussed previously in the literature \cite{SinghCederBazant112008}. 

\begin{figure}
\centering{}\includegraphics[width=0.7\linewidth]{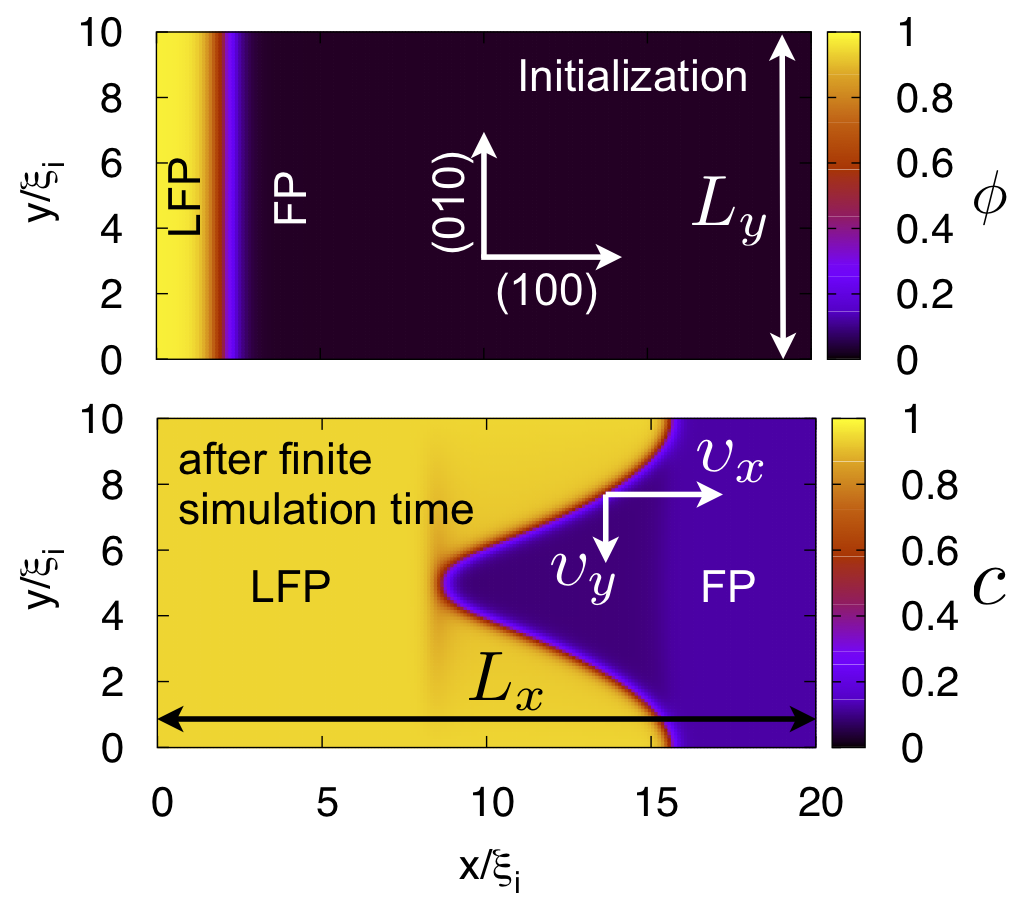} \caption{\label{fig:sim-2D-setup} (color online) Setup of the 2D phase field simulations of the lithiation process
of a rectangular LiFePO\textsubscript{4}-particle. On top we show the initial phase field $\phi$, and
below the Li-ion concentration field after finite simulation time is shown. (LFP denotes the LiFePO\textsubscript{4}-
phase and FP the FePO\textsubscript{4}-phase)}
\end{figure}

The rectangular 2D domain of size $L_{x}\times L_{y}$ represents a 2D cut of a partially lithiated particle
parallel to the {[}001{]}-crystallographic plane. The rectangular domain is oriented such that the $x-$
and $y-$coordinate axis meet with the other crystallographic directions $(100)$ and $(010)$, respectively,
as shown in Fig.~\ref{fig:sim-2D-setup}. Due to the olivine crystal structure, the lithium transport
is limited to one dimensional channels along the $(010)$ direction, and it remains only a single non-vanishing
element in the mobility tensor, i.e.~the $M_{yy}-$element. Furthermore, the domain contains finite
regions of LFP- and FP-phase, being well separated by a diffuse interface of width $2\xi_{i}$. The lithiation
of the 2D-cut of the particle is related to a phase transformation from the lithium poor FP-phase to
the lithium rich LFP-phase. The lithiation of the particle is driven by a positive non equilibrium chemical
potential $\mu_{0}$ applied on the top and bottom boundary of the 2D simulation, which enforces a finite
lithium influx into the particle favoring the transformation from FP-phase to LFP.

We start from the FP-phase with a small strip of LFP phase placed on the left side of the system, as
shown in Fig.~\ref{fig:sim-2D-setup}. After a certain transient time the system reaches a steady state,
where the FP/LFP-interface is moving with a constant velocity and constant shape. For $L_{y}\gg\xi_{i}$,
the interface exhibits a parabolic shaped stead state profile, because the solid solid phase transformations
is limited by the one dimensional lithium diffusion along the crystallographic $(010)-$direction (see
Fig.~\ref{fig:sim-2D-setup}). Similar parabolic shaped LFP/FP interfaces have also been reported in
other phase-field simulation research-work by Tang et al.~\cite{TangBelakDorr032011}, who employed
a Cahn-Hilliard-type formulation there.

\begin{figure}
\centering{}\includegraphics[viewport=0bp 0bp 360bp 252bp,width=0.7\columnwidth]{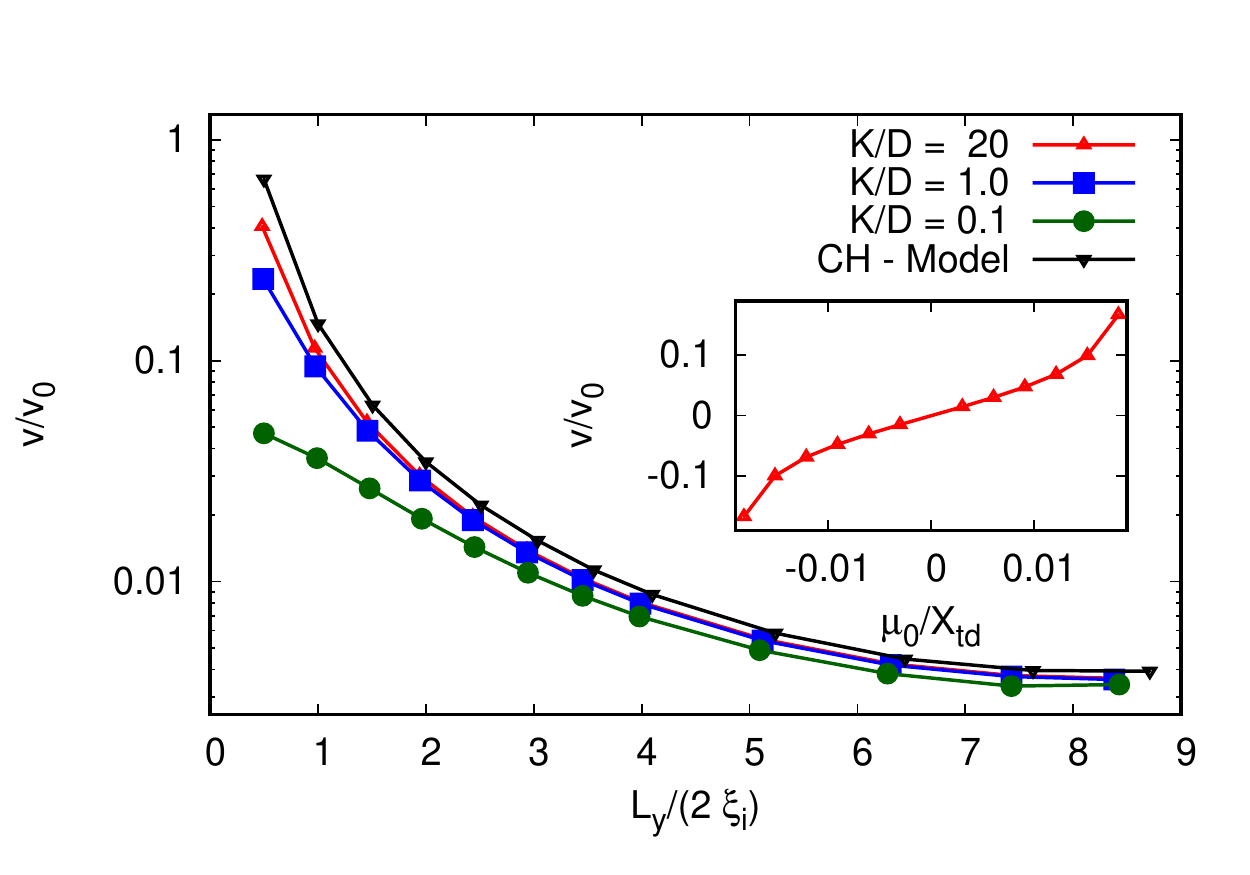}\caption{\label{fig:v_0} (color online) Change of the steady state interface velocity as function of the particle
height $L_{y}$. The velocity is normalized to $v_{0}=D/\xi_{i}$. Inset: Dimensionless interface velocity
measured under different external potentials $\mu_{0}/X_{td}$. }
\end{figure}

During the simulation, we measure the velocity $v_{x}$ of the moving solid solid interface. The resulting
steady state velocity is selected by the Li-ion influx through the top and bottom boundary, and is controlled
by the imposed external chemical potential $\mu_{0}$. If $\mu_{0}$ is chosen to be positive, then Li
is transported into the particle, and the LFP-phase starts to grow at the expense of the FP-phase. For
$\mu_{0}\:<\:0$ Li-ions are extracted leading to the growth of the FP phase. In the inset of Fig.~\ref{fig:v_0},
we plot the measured steady state velocity $v_{0}$ as a function of the external potential $\mu_{0}$,
for a ratio of the system hight to the double of the phase field width of $L_{y}/2\xi_{i}=1$. The dimensionless
ratio of the thermodynamic factor times the interface width to surface energy has been chosen to be $X_{td}\xi_{i}/\gamma_{i}=130$.
The velocity is normalized to the characteristic velocity $v_{0}=D/\xi_{i}$.

In the main plot of Fig.~\ref{fig:v_0}, we show the steady state velocity of the LFP/FP-phase boundary
as function of the particle height $L_{y}$, for a ratio of the external chemical potential to the thermodynamic
factor of $\mu_{0}/X_{td}=0.01$. For the dimensionless ratio of the thermodynamic factor times the interface
width to surface energy, we again impose $X_{td}\xi_{i}/\gamma_{i}=130$. Generally, we observe an increasing
steady state velocity for a decreasing particle-height. And for particles much higher than the width
of the diffuse interface region $\xi_{i}$, the velocity becomes independent from the particle height
$L_{y}$. Upon a decrease of the dimensionless height from 8 to 1 the Cahn-Hilliard-model predicts a
drastic increase of the steady state velocity by nearly two orders of magnitude. Within the Allen-Cahn
formulation, a similar velocity-increase is observed, when the phase-field kinetic coefficient $K$ is
chosen to be equally or larger than the Li-diffusion coefficient $D$, i.e.~$K/D\geq1$. When the imposed
value for the kinetic coefficient $K$ is a factor of ten lower than that of $D$, the respective increase
of the steady state transformation velocity with decreasing particle height turns out to be respectively
smaller. At the same time, the limiting value for the steady state velocity in particles much higher
than the phase-field width is found to be nearly independent from the imposed ratio $K/D.$ 

We point out that such an increase of the transformation velocity with a decreasing particle height,
is intrinsically related to the underlying diffuse interface description. It has already been reported
earlier by Singh et.~al, who also did computational studies using a Cahn-Hilliard type phase-field model
\cite{SinghCederBazant112008}. However, as compared to the conventional Cahn-Hilliard type description,
our Allen-Cahn-type model further allows to control the strength of this velocity-increase, by the respective
adjustment of the additional kinetic parameter $K$, as shown in the main plot of Fig.~\ref{fig:v_0}. 

\subsection{LFP to FP transformations including elastic effects}

\begin{table}
\caption{\label{tab:The-elastic-parameters}The elastic parameters used in the simulations of the lithiation of
a rectangular particle in three dimensions. }

\centering{}%
\begin{tabular*}{1\textwidth}{@{\extracolsep{\fill}}>{\raggedright}p{0.1\textwidth}rlllll>{\raggedright}p{0.18\textwidth}}
\hline 
\noalign{\vskip5pt}
\multicolumn{2}{r}{Parameter} &  &  LiFePO\textsubscript{4} & FePO\textsubscript{4} & Electrolyte & Dimension & Literature\tabularnewline[4pt]
\hline 
\noalign{\vskip3pt}
\multirow{9}{0.1\textwidth}{Stiffness tensor } & $C_{11}$ &  & $136$  & $171.2$  & $0$  & $\mathrm{GPa}$ & \multirow{9}{0.18\textwidth}{DFT-calculations (average of $GGA$ and $GGA+U$) \cite{MaxischCeder052006} }\tabularnewline
\noalign{\vskip3pt}
 & $C_{22}$ &  & $200$  & $141$ & $0$  & $\mathrm{GPa}$ & \tabularnewline
\noalign{\vskip3pt}
 & $C_{33}$ &  & $173$  & $128$ & $0$  & $\mathrm{GPa}$ & \tabularnewline
\noalign{\vskip3pt}
 & $C_{44}$ &  & $35.9$  & $35.7$ & $0$  & $\mathrm{GPa}$ & \tabularnewline
\noalign{\vskip3pt}
 & $C_{55}$ &  & $49.2$  & $45.3$ & $0$  & $\mathrm{GPa}$ & \tabularnewline
\noalign{\vskip3pt}
 & $C_{66}$ &  & $45.0$  & $50.6$ & $0$  & $\mathrm{GPa}$ & \tabularnewline
\noalign{\vskip3pt}
 & $C_{12}$ &  & $73.6$  & $31.3$ & $0$  & $\mathrm{GPa}$ & \tabularnewline
\noalign{\vskip3pt}
 & $C_{13}$ &  & $53.4$  & $55.6$ & $0$  & $\mathrm{GPa}$ & \tabularnewline
\noalign{\vskip3pt}
 & $C_{23}$ &  & $50.5$  & $14.4$ & $0$  & $\mathrm{GPa}$ & \tabularnewline
\hline 
\noalign{\vskip3pt}
\multirow{4}{0.1\textwidth}{Misfit strain } & $\epsilon_{1}^{0}$ &  & $+2.5\cdot10^{-2}$  & $-2.5\cdot10^{-2}$ & $0$  & - & \multirow{4}{0.18\textwidth}{XRD-measurements \cite{MeethonHuangCarter032007,TangBelakDorr032011}}\tabularnewline
\noalign{\vskip3pt}
 & $\epsilon_{2}^{0}$ &  & $+1.8\cdot10^{-2}$  & $-1.8\cdot10^{-2}$ & $0$  & - & \tabularnewline
\noalign{\vskip3pt}
 & $\epsilon_{3}^{0}$ &  & $-0.85\cdot10^{-2}$  & $+0.85\cdot10^{-2}$ & $0$  & - & \tabularnewline
\noalign{\vskip3pt}
 & $\epsilon_{4-6}^{0}$ &  & $0$  & $0$  & $0$  & - & \tabularnewline
\hline 
\end{tabular*}
\end{table}

\begin{figure}
\centering{}a)\includegraphics[width=0.28\columnwidth]{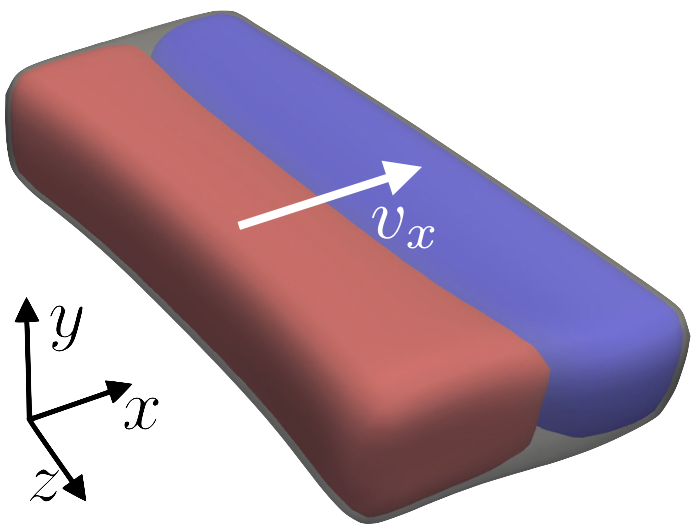} ~ b)\includegraphics[width=0.28\columnwidth]{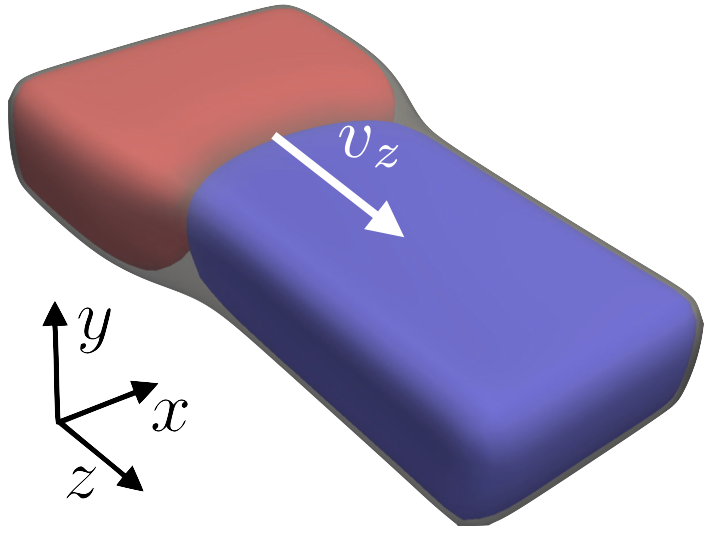}
~ c)\includegraphics[width=0.3\columnwidth]{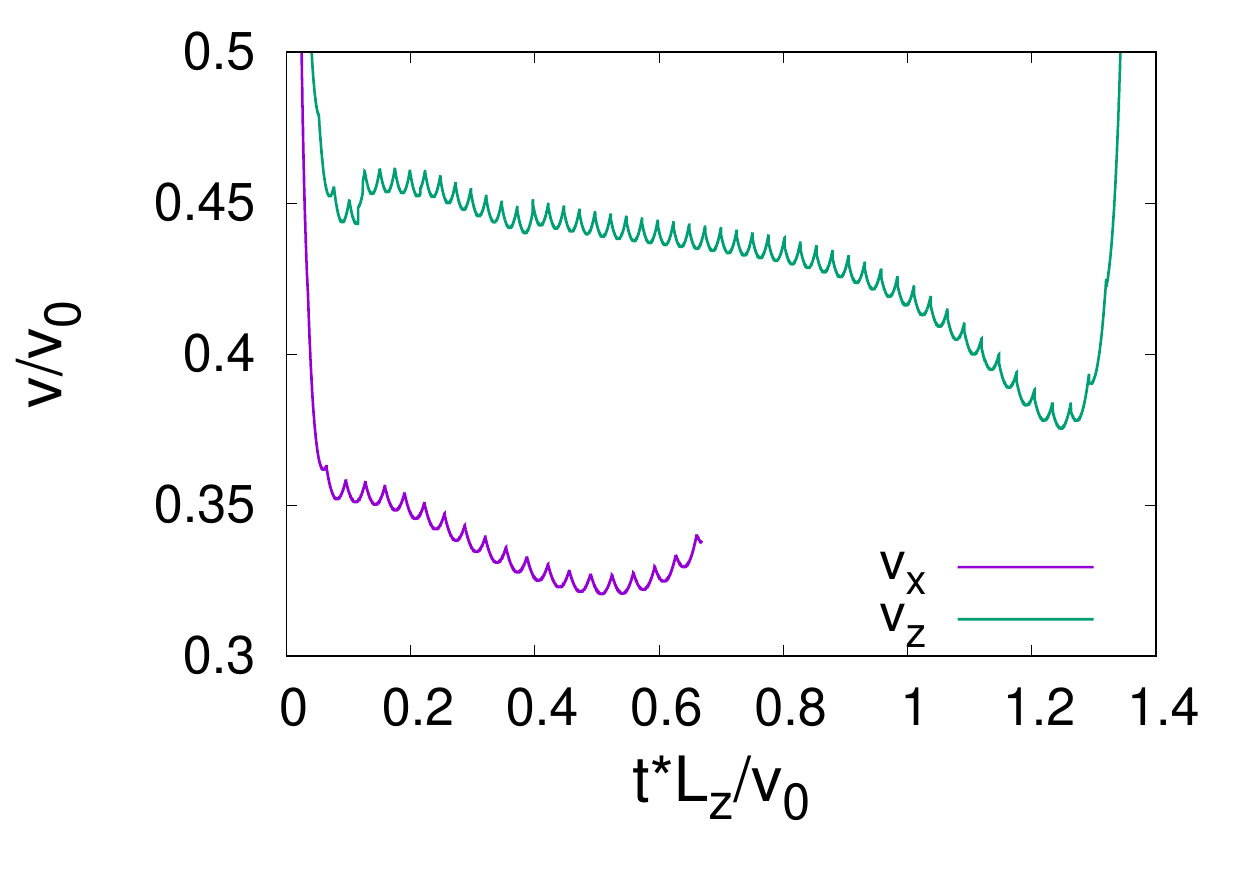}\caption{\label{fig:3D-velocity-configurations} (color online) Snapshots of the different 3D lithiation configurations.
a) LFP/FP phase-front is growing along $x-$direction. b) LFP/FP phase-front is growing along $z$-direction.
The elastic deformations due to the anisotropic lattice misfit have been magnified by a factor 10 for
better visualization. c) Plot of the momentary interface velocity as measured during the two different
simulation. The velocity is normalized to $v_{0}=D/\xi$. }
\end{figure}

Next, the full model is applied to the lithiation-kinetics of a rectangularly shaped particle surrounded
with a thin layer of the soft electrolyte phase in three dimensions. Here, we also include the effects
from the anisotropic and inhomogeneous elastic deformations, which result from the lattice misfit in
case of coherent connection between the LFP and the FP phase. The elastic parameters used in the respective
simulations are listed in Tab.~\ref{tab:The-elastic-parameters}. Here, the anisotropic misfit strain
reflects the crystallographic shape-change of a unit-volume during the transformation from the orthorhombic
FePO\textsubscript{4}-phase to the also orthorhombic LiFePO\textsubscript{4}-phase. Here, we distribute
the misfit strain symmetrically onto the two different solid phases. It is calculated from the three
different lattice parameters, as follows 
\begin{align*}
\left(\epsilon_{ii}^{0}\right)^{LFP} & =\frac{a_{i}^{LFP}-a_{i}^{FP}}{a_{i}^{LFP}+a_{i}^{FP}},\\
\left(\epsilon_{ii}^{0}\right)^{FP} & =\frac{a_{i}^{FP}-a_{i}^{LFP}}{a_{i}^{LFP}+a_{i}^{FP}},
\end{align*}
where the index $i$ denotes one of the three different basic crystallographic orientations. Note, that
this definition of the different eigenstrains attributed to the two different solid phases, implies that
the elastic reference state corresponds to the arithmetic average of the FP and the LFP. From a physical,
important is mainly the different in the eigenstrains of the two phases. Other dimensionless parameters
used in these two simulations are $K/D=20$, $\mu_{0}/X_{td}=0.1$, $X_{td}\xi_{i}/\gamma_{i}=6$ and
$X_{td}/C_{22}^{LFP}=5$. The considered configurations are provided in the Figs.~\ref{fig:3D-velocity-configurations}a)
and b), where snapshots of the respective simulations are shown. 

In order to visualize the presence of the elastic effects, respectively deformed $0.5$-contours of the
LFP-phase (red), the FP-phase (blue) as well as the electrolyte-phase (transparent gray) have been plotted.
Note that for better visibility the elastic deformations have been magnified by a factor 10. In Fig.~\ref{fig:3D-velocity-configurations}c),
the momentary growth-velocity of the LFP/FP-interface, growing along the $x-$direction case a), is compared
to the case b), when it grows along the $z-$direction. The observed difference in the lithiation-kinetics
is related to the influence from the effects the anisotropic and inhomogeneous elasticity, which result
from the lattice misfit. 

\subsection{Formation of the anisotropic particle-shape}

Further, the lithiation or delithiation kinetics of a single LiFePO\textsubscript{4}-particle depends
on the particle's size and shape in at least two different respects: Coherency strains and stresses as
well as diffusion paths are both shape and size dependent. Therefore, a central goal of the modeling
has been to include realistic shapes into the model. Experimentally quite a number of different particle
shapes are known for the LiFePO\textsubscript{4}. For example, crystallites have been observed in the
shape of hexagonal- or diamond-type platelets \cite{ChenSongRichard042006,QuiWangXieLiWenWang2012},
rectangular prisms\cite{EllisWang2007}, rods and block shapes \cite{FisherIslam2008}. Especially the
hexagonal- or diamond-type platelets seem to be quite favorable since the due to their large $(010)$-oriented
surface. This surface is normal to the most facile pathway for lithium ion migration, and is hence probably
the electrochemically most active one. Further, the thinner the particle in the $(010)$ direction is
the shorter are the lithium diffusion paths inside the particle. This may enhance the rate capability
of a cathode prepared from such a material \cite{FisherIslam2008}.

 Here, the aim is to generate a strongly anisotropic shape such as the diamond-type platelets-shape,
which has been experimentally observed in hydro-thermally synthesized LiFePO\textsubscript{4} \cite{ChenSongRichard042006}.
Unfortunately, such a strongly anisotropic shape can not be produced using Ab-initio-based surface energies.
The respectively surface energies for different interface orientations are to similar in value to produce
the experimentally measured aspect-ratios of 3 to 10 \cite{WangZhouMengCeder2007,FisherIslam2008}. Therefore,
using the anisotropy functions given in Eqs.~(\ref{eq:Ani-Function-alpha}) and (\ref{eq:Ani-Function-beta}),
an anisotropy profile including some experimentally observed facet-orientations has been designed such
that is visually fits to the experimentally observes particle shapes \cite{ChenSongRichard042006}. The
anisotropy parameters used in the subsequent phase-field simulation for the particle generation are given
in Tab.~\ref{tab:The-anisotropy-parameters}. 
\begin{table}
\caption{\label{tab:The-anisotropy-parameters}The anisotropy parameters used to generate the anisotropic LiFePO\textsubscript{4}-particle
shown in Fig.~\ref{fig:par_3D} }

\centering{}%
\begin{tabular*}{1\textwidth}{@{\extracolsep{\fill}}rrlcc}
\hline 
\noalign{\vskip5pt}
Orientation &  & Function & Exponent $d$ & Amplitude $\delta$\tabularnewline[4pt]
\hline 
\noalign{\vskip3pt}
$(010)$ &  & $\beta(\nabla\varphi)$; Eq.~(\ref{eq:Ani-Function-beta}) & - & $5.0$ \tabularnewline
\noalign{\vskip3pt}
$(100)$ &  & $\alpha(\nabla\varphi)$; Eq.~(\ref{eq:Ani-Function-alpha}) & 0 & $3.0$ \tabularnewline
\noalign{\vskip3pt}
$(100)$ &  & $\alpha(\nabla\varphi)$; Eq.~(\ref{eq:Ani-Function-alpha}) & 5 & $0.08$ \tabularnewline
\noalign{\vskip3pt}
$(101)$, $(\underline{1}01)$ &  & $\alpha(\nabla\varphi)$; Eq.~(\ref{eq:Ani-Function-alpha}) & 5 & $0.24$ \tabularnewline
\noalign{\vskip3pt}
$(201)$, $(\underline{2}01)$ &  & $\alpha(\nabla\varphi)$; Eq.~(\ref{eq:Ani-Function-alpha}) & 5 & $0.15$ \tabularnewline
\hline 
\end{tabular*}
\end{table}
The aim is to get an anisotropic equilibrium shape having an aspect ratio of 1/5 in the yz-cut and having
a respectively dominant facet in the $(010)$-direction. Therefore, first a respectively strong facet
has been set in that direction using the anisotropy function from Eq.~(\ref{eq:Ani-Function-beta}),
see the first line of Tab.~\ref{tab:The-anisotropy-parameters}. Then, within the xz-cut of the resulting
particles, we want an aspect ratio of 1/2, which requires the additional overlay of a smooth anisotropy
profile with the orientation $(100)$, such as given by line two of the table. The following lines in
Tab.~\ref{tab:The-anisotropy-parameters} each provide respective facets at the given orientations respectively
on the front and backside of the particle.

\begin{figure*}
\begin{centering}
a)\includegraphics[width=0.45\linewidth]{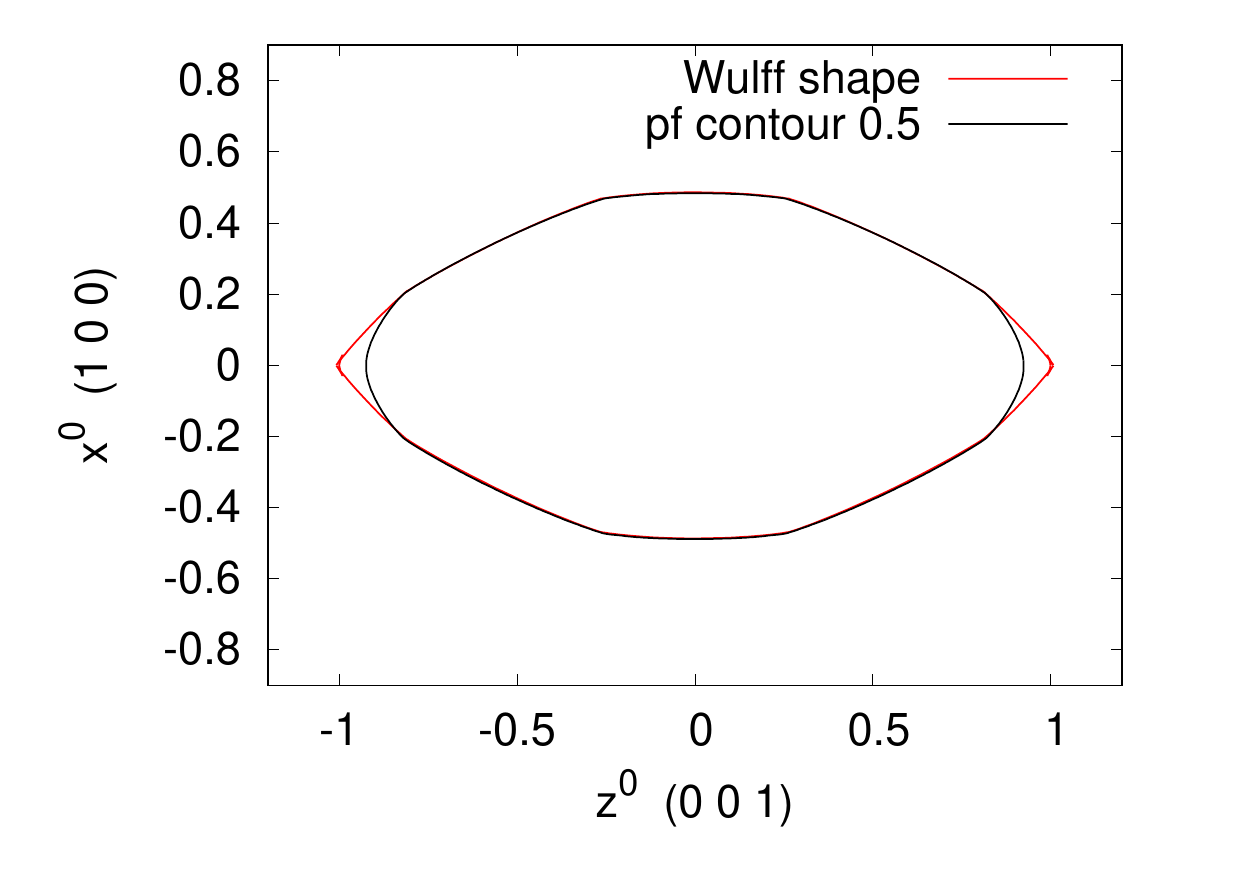}~b)\includegraphics[width=0.45\linewidth]{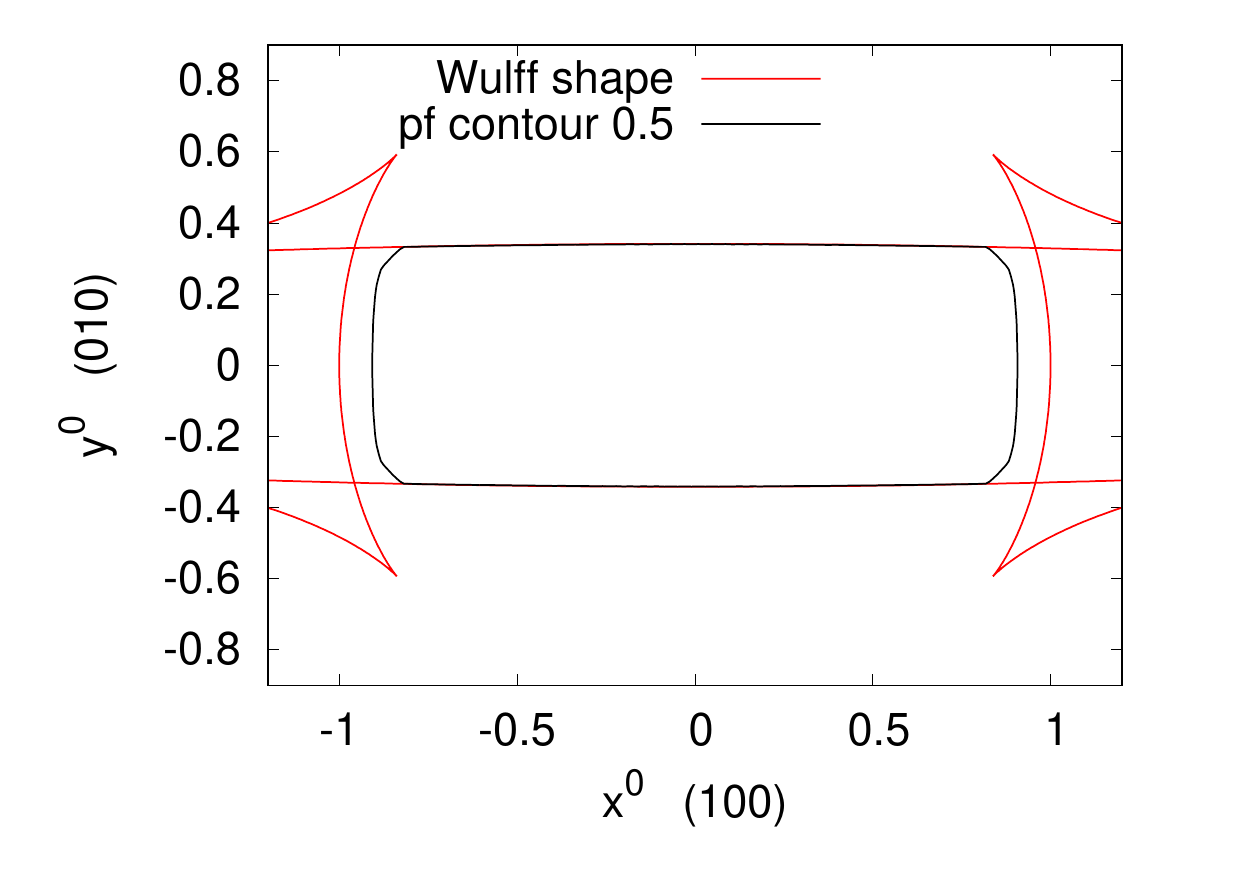}
\par\end{centering}
\begin{centering}
c)\includegraphics[width=0.7\linewidth]{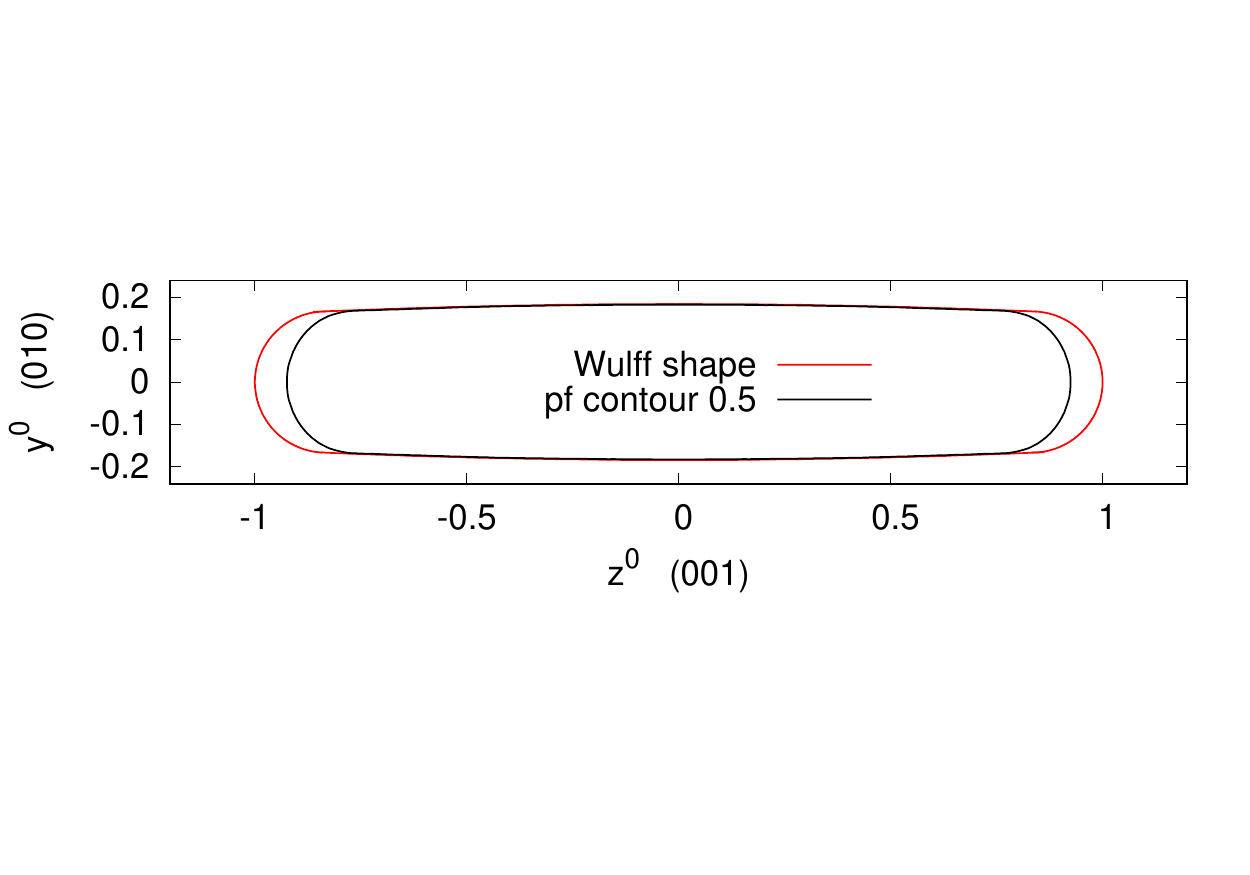} 
\par\end{centering}
\caption{\label{fig:2D-Wulff-shapes} (color online) Comparison of the analytic Wulff-shape with the $\varphi=1/2-$contour
of the xz-cut of the relaxed anisotropic 3D phase field, discriminating between the inner and outer of
the LiFePO\textsubscript{4}-particle. In a) the xz-cut through the particles mass center is shown, in
b) the respective xy-cut and in c) the yz-cut. }
\end{figure*}

In Fig.~\ref{fig:2D-Wulff-shapes}, we compare the resulting 2D cuts of the simulated 3D phase-field
with the respective 2D Wulff-construction. For a given anisotropy profile the equilibrium shape is given
by Wulff's theorem. The later demands that the particle shape is given by the multitude of all tangents
of the $\gamma_{p}(\nabla\varphi)-$plot \cite{BurtonCabreraFrank061951}. Thus, in 2D the equilibrium
shape of a particle in parametric form is given by 
\begin{align}
x= & \gamma(\vartheta)\cos(\vartheta)-\gamma'(\vartheta)\sin(\vartheta),\label{eq:Eq-Shape-Wulff-x}\\
y= & \gamma(\vartheta)\sin(\vartheta)+\gamma'(\vartheta)\cos(\vartheta),\label{eq:Eq-Shape-Wulff-y}
\end{align}
where $\vartheta$ is the polar angle. For sufficiently strong anisotropies, the particle equilibrium
shape develops sharp corners, where certain high energy orientations are excluded. This is accompanied
by the appearance of the ``ears'', i.e.~metastable and unstable branches, in the Wulff-construction,
as can be also seen in the figure. With regard to the anisotropy profile chosen here, these ``ears''
are very dominate in the xy-cut, as shown in Fig.~\ref{fig:2D-Wulff-shapes}b). Generally, with regard
to the specific choice of the anisotropy profile, the appearance of ears should be avoided as much as
possible, as the phase-field evolution is ill-posed at the respective locations \cite{TaylorCahn021998,EgglestMcFaddeVoorhee032001,FleckMushongPilipen102011}.

\section{Conclusion}

\label{sec:summary}In summary, we have developed a continuum phase-field model, which takes into account
a whole range of mechanisms relevant in the lithiation-reaction in LiFePO\textsubscript{4} cathode materials,
explicitly resolving the particle with it's anisotropic shape: 
\begin{itemize}
\item The introduction of anisotropic bulk diffusion along 1D channels in (010) crystallographic direction 
\item The incorporation of anisotropic coherency strains arising from the lattice-mismatch between the two
joining solid phases with different elastic constants. 
\item The implementation of strongly anisotropic interfacial energies that give rise to the anisotropic particle-shapes 
\item The introduction of the anisotropic particle-shape, which act as a free-surface guaranteeing realistic
strain energy contributions 
\end{itemize}
A major difference of the presented model, as compared to other models of similar purpose is that the
original Cahn-Hilliard-problem was reformulated in terms of two strongly coupled but still independent
kinetic equations (diffusion, solid-solid phase transformations). This lead to two coupled 2nd order
partial differential equations of parabolic type, instead of one equation of forth order, which is numerically
advantageous and leads to the development of kinetic depletion-zones which we regard as physical, and
which are not present in the Cahn-Hilliard-type formulation.

Moreover, first interesting results on the size dependent kinetics of diffusion limited phase transformations
are presented. The respective simulation study was restricted to a rectangular 2D domain representing
a 2D cut of a partially lithiated particle along the $(001)$-plane and neglecting elastic effects. Then
the transformation kinetics is controlled by the external chemical potential $\mu_{0}$, which acts as
a driving force for the diffusion limited transformation wave. In the range of small particle thickness
in the $L_{y}$ direction the simulation studies indicate that $L_{y}$ has a strong impact onto the
Li-transport into the bulk. In the case of diffusion limited transformations, we observe a strong increase
of the resulting steady state velocity, which is related to the difference in the two kinetic time-scales

Finally, we discuss the generation of physically motivated anisotropic particle shapes. From studying
experimental particle shapes, we picked out a number of interesting facet-orientations. Then an anisotropy
profile was set out for these facet-orientations, and we tuned the respectively calculated Wullf-construction
such that is optically fitted to the experimentally observes particle shapes. The resulting anisotropy
parameters have been used in the subsequent 3D phase-field simulation for the particle generation. Contours
of 2D sections of the resulting phase-field, have been compared to the original 2D Wulff-constructions.
This study demonstrates that also the simulation of strongly anisotropic particles with aspect ratios
up to 1/5 is possible. Special care, has to be given to the setup of the anisotropy profile, such that
the appearance of ears in the Wulff-shape should be avoided as much as possible.
\begin{acknowledgments}
We thank the Federal Ministry for Economics and Energy (BMWi) of the Federal Republic of Germany for
the financial support under the running project COORETEC: ISar (funding code: 03ET7047D). Further, we
thank the german research foundation for financial support under the DGF Priority Program SPP 1473/1.
We acknowledge the financial support from the Oberfranken-Stiftung and the state of Bavaria of the Federal
Republic of Germany. Finally, we thank Prof.~F.~Mertens for the fruitful discussions on the thermodynamic
formulation of the model. 
\end{acknowledgments}

\section*{Data availability}

The raw data as well as the processed data required to reproduce these findings are provided within the
above text (see for instance Tab.~\ref{tab:The-elastic-parameters} and Tab.~\ref{tab:The-anisotropy-parameters}).

\bibliographystyle{mynaturemag}
\bibliography{Li-BatteriesSPP1473}

\end{document}